\documentclass[12pt,reqno]{article}
\usepackage{geometry}                % See geometry.pdf to learn the layout options. There are lots.
\usepackage{graphicx}
\usepackage{amssymb}
\usepackage{epstopdf}
\usepackage{subfigure}
\usepackage{setspace}
\usepackage{amsmath,amsthm}
\numberwithin{equation}{section}
\newcommand{\R}{\ensuremath{\mathbb{R}}}

\newcommand{\dxo}{\ensuremath{\delta x(0)}}
\newcommand{\da}{\ensuremath{\delta \alpha}}
\newcommand{\dx}{\ensuremath{\delta x}}
\newcommand{\dxx}{\ensuremath{\delta^2 x}}
\newcommand{\T}{\ensuremath{\mathsf{T}}}
\usepackage{fancyhdr} 
\fancypagestyle{plain}{} 

\begin{document}
\begin{titlepage}
\rhead{ }
\cfoot{ }
\lfoot{*Corresponding Author: varahan@ou.edu}
\title{Forecast Bias Correction: A Second-Order Method}
\author{Sean Crowell\\Department of Mathematics\\University of Oklahoma\\Norman, OK 73019\\scrowell@ou.edu \and S. Lakshmivarahan*\\School of Computer Science\\University of Oklahoma\\Norman, OK 73019\\varahan@ou.edu}
\date{ }
\maketitle

\begin{abstract}
In this paper we describe a methodology for determining corrections to parameters and initial conditions in order to improve model forecasts in the presence of data.  This methodology is grounded in the variational problem of minimizing the norm of the errors between forecast and data.  The general method is presented, and then the method is applied to the specific example of a scalar model, the logistic equation, where it is shown that the method produces satisfactory results.
\end{abstract}
\vfill
Acknowledgement:  We wish to record our sincere thanks to John Lewis for his interest in this work.\\
\\
\center
\today
\vfill
\end{titlepage}
\lfoot{ } 
\cfoot{ \thepage }
\rhead{ }
\lhead{ }
\doublespace
\section{Introduction}
Models are abstractions of reality and the goodness of a model is often judged by its ability to explain the observations which are the reflections of the underlying reality.  The difference between the actual observation and the model counterpart of the observation is known as the \textit{prediction error} or \textit{forecast bias}.  This error/bias is a result of either an "inadequate" model or a result of misspecifications of initial/boundary conditions and/or parameters in the model.  Recently, Lakshmivarahan and Lewis (2008) (LL(2008), hereafter) have developed a framework for estimating the error in initial/boundary conditions and/or parameters that will account for the observed forecast bias.  Using the (first-order) sensitivity of the solution with respect to the initial conditions and parameters and the observed forecast errors, this framework recasts the bias estimation problem as an inverse problem, in particular as a linear least-squares problem.  In LL(2008), they demonstrated the power of this idea using a simplified model consisting of a system of three coupled nonlinear ordinary differential equations that describes the evolution of the mixed-layer over the Gulf of Mexico and actual observations obtained from an earlier field experiment.  For details, refer to LL(2008) and references therein.

In this paper we further analyze the power of this framework using a scalar model for population growth known as the logistic model in continuous time.  But instead of the first-order method used in LL(2008), in this paper we compare the performance of first-order and second-order methods.  For simplicity it is assumed that the model is adequate and the forecast errors are due only to errors in the initial condition and the parameter.  As will become evident, the first-order method leads to a linear least-squares problem and the second-order method leads to a nonlinear least-squares problem.  Analysis of model inadequacies will be explored in a companion paper.

Section 2 contains a summary of the framework and is adapted from LL(2008).  Properties of the logistic model are described in Section 3.  Numerical experiments related to the estimation of errors using the first-order and second-order methods along with a comparison of these methods are contained in Section 4.  Concluding observations are contained in Section 5.

\section{The Framework - a Summary}
Let $x(t) \in \R$ denote the state of a dynamical system at time $t \in \R^+$ where $x(0)$ is called the initial state.  Let $f : \R \mathsf{x} \R \mathsf{x} \R^+ \rightarrow \R$ and the state evolve according to a scalar nonlinear ordinary differential equation given by
\begin{equation}\label{model}
\dot{x} = f(x,\alpha,t)
\end{equation}
where $\alpha \in \R$ is called the parameter.  It is tacitly assumed that the function $f$ in \eqref{model} satisfies all of the conditions required for the existence, uniqueness, and smoothness of the solution $x(t) = x(t,\alpha,x(0))$ for all $t \in \R^+$.

Let $h: \R \rightarrow \R$ be a smooth function that defines the model counterpart of the observation
\begin{equation}\label{obsop}
z_t^M = h(x(t))
\end{equation}

Let $z_t$ be the actual observation obtained from the field measurement.  Then
\begin{equation}\label{error}
e_t = z_t-z_t^M
\end{equation}
denotes the prediction error or the forecast bias.  It is assumed that the model in \eqref{model} does not have any deficiencies and the forecast bias is mainly due to the misspecification of the initial condition, $x(0)$, and/or the parameter $\alpha$.

Let $\Delta x$ be the actual change in $x(t)$ induced by the perturbations $\dxo$ in $x(0)$ and $\da$ in $\alpha$.  The goal is to find $\beta = (\dxo, \da)^\T$, the vector of perturbations such that 
\begin{equation}
z_t-h(x_t+\Delta x) = 0.
\end{equation}
Let $\Delta h$ be the actual change in $h(x(t))$ induced by the change $\Delta x$ in $x(t)$.  Then
\begin{equation}\label{obsoppert}
h(x(t)+\Delta x) = h(x(t))+\Delta h
\end{equation}
Combining \eqref{error} - \eqref{obsoppert}, our goal is to find $\beta$ such that 
\begin{equation}\label{dherr}
e_t = \Delta h
\end{equation}
Recall that 
\begin{equation}\label{htaylor}
\Delta h = \sum_{j=1}^\infty \delta^j h
\end{equation}
where $\delta^j h$ is called the $j^th$ variation of $h$, the fraction of the induced change that is attributable to the $j^{th}$ derivative of $h$ and the change in $\Delta x$ in $x(t)$.  Similarly
\begin{equation}\label{xtaylor}
\Delta x = \sum_{j=1}^\infty \delta^j x
\end{equation}
where $\delta^jx$, called the $j^{th}$ variation of $x$, is the fraction of the total change in $x(t)$ that is attributable to the $j^{th}$ partial derivatives of $x(t)$ with respect to $\alpha$ and $x(0)$ and $\beta$.

In practice we approximate the infinite sum by taking only the first $k$ terms, resulting in a $k^{th}$-order approximation given by
\begin{equation}\label{hxtaylor}
\Delta h = \sum_{j=1}^k \delta^j h \mbox{ and } \Delta x = \sum_{j=1}^k \delta^j x.
\end{equation}

\subsection{First-order Approximation}
Setting $k = 1$, we get
\begin{equation}\label{dhdx}
\Delta h = \delta h \mbox{ and } \Delta x = \dx.
\end{equation}
Let $D_a(g)=\frac{\partial g}{\partial a}$.  From first principles and \eqref{dhdx}, it follows that
\begin{equation}\label{derh}
\delta h = D_x(h)\Delta x = D_x(h)\dx
\end{equation}
and
\begin{equation}\label{derx}
\dx = D_{x(0)}(x)\dxo + D_{\alpha}(x)\da.
\end{equation}
The first derivative $D_a(g)$ is called the first-order sensitivity of $g$ with respect to $a$ (Cacuci (2003) and Cruz(1973)).  Setting
\begin{equation*}\label{matrix}
\overline{H_1}=D_{x(0)}(x), \mbox{ } \overline{H_2} = D_{\alpha}(x), \overline{H} = [\overline{H_1},\overline{H_2}] \in \R^{1\mathsf{x}2},
\end{equation*}
\eqref{derx} can be rewritten as
\begin{equation}\label{derxmatrix}
	\dx = \overline{H}\beta.
\end{equation}
Combining \eqref{derxmatrix} and \eqref{derh} with \eqref{dherr} we obtain an underdetermined linear least-squares problem
\begin{equation}\label{hbe}
	H\beta = e
\end{equation}
where
\begin{equation}\label{h=dhhbar}
	H = D_x(h)\overline{H} \in \R^{1 \mathsf{x} 2}
\end{equation}
Suppose that there are $N$ observations $z_{t_1}$, $z_{t_2}$, ... $z_{t_N}$ available at times $0 \leq t_1 < t_2 < ... < t_N$.  Then, at each time $t_i$, $1 \leq i \leq N$, we have the forecast error
\begin{equation}\label{ei}
	e_{t_i} = z_{t_i} - h(x_{t_i})
\end{equation}
Define
\begin{align}
	\overline{H_1}({t_i}) &= D_{x(0)}(x(t_i)), \mbox{ } \overline{H_2}(t_i) = D_{\alpha}(x(t_i)) \label{hi} \\
	\overline{H}({t_i}) &= [\overline{H_1}(t_i), \overline{H_2}(t_i)] \in \R^{1\mathsf{x}2} \notag
\end{align}
and a diagonal matrix
\begin{equation*}
	D(h) = \text{Diag}\{D_1(h),D_2(h),...,D_N(h)\} \in \R^{N\mathsf{x}N}
\end{equation*}
where
\begin{equation*}
D_i(h) = D_x(h(x(t_i))).
\end{equation*}
for simplicity in notation.  Let
\begin{equation}\label{hbarn}
	\overline{H_N}=\left[ \begin{matrix} 
	\overline{H_1}(t_1) & \overline{H_2}(t_1)\\
	\overline{H_1}(t_1) & \overline{H_2}(t_1) \\
	\vdots & \vdots \\
	\overline{H_1}(t_N) & \overline{H_2}(t_N) \\
	\end{matrix} \right] \in \R^{N\mathsf{x}2}
\end{equation}
\begin{equation}\label{hn}
	H_N = D(h)\overline{H_N}
\end{equation}
and
\begin{equation}\label{en}
	E_N=(e_{t_1}, e_{t_2},...,e_{t_N})^\T \in \R^N
\end{equation}
For this case of $N$ observations, in place of \eqref{hbe}, we obtain a linear least-squares problem
\begin{equation}\label{hnb=en}
	H_N\beta = E_N
\end{equation}
where $H_N \in \R^{N\mathsf{x}2}$, $\beta \in \R^2$, and $E_N \in \R^N$.

The unknown $\beta$ is obtained by minimizing
\begin{equation}\label{lsfunctional}
	g_1(\beta) = \|H_N\beta-E_N\|^2
\end{equation}
where it is tacitly assumed that the matrix $H_N$ is of full rank.  Under this assumption, $\beta$ is given by (Lewis, et al (2006))
\begin{equation}\label{sol}
	\beta = \left\lbrace \begin{matrix} 
	\left(H_N^\T H_N\right)^{-1}H_N^\T E_N & \text{if } N > 2 \\
	H_N^{-1} E_N & \text{if } N = 2 \\
	H_N^\T \left(H_N H_N^\T \right)^{-1}E_N & \text{if } N < 2 \\
	\end{matrix} \right. 
\end{equation}
When $H_N$ is not of full rank, we invoke the Tikhonov regularization (Lewis, et al (2006)) using which $\beta$ is obtained by minimizing
\begin{equation}\label{tik}
	\overline{g_1}(\beta) = \| H_N\beta - E_N \|^2 +\frac{\lambda^2}{2}\|\beta\|^2
\end{equation}
for some real constant $\lambda > 0$,  called the regularization parameter.  The minimizing $\beta$ in this case is given by 
\begin{equation}\label{betatik}
	\overline{\beta}_{LS} = (H_N^\T H_N + \lambda I)^{-1}H_N^\T E_N
\end{equation}
In place of Tikhonov regularization, one could use the generalized inverse of the matrix $H_N$ to obtain
\begin{equation*}
\beta = H_N^+ E_N
\end{equation*}
where $H_N^+$ is the Moore-Penrose inverse (Lewis et al (2006)).
\subsection{Second-Order Approximation}
Setting $k = 2$, in \eqref{hxtaylor}, it follows that
\begin{equation}\label{dxx}
	\Delta x = \dx+\dxx	
\end{equation}
where $\dx$ is given by \eqref{derxmatrix} is linear in $\beta$.  From first principles, we obtain
\begin{equation}\label{d2x}
	\dxx = \frac{1}{2}\beta^\T D^2(x) \beta
\end{equation}
which is quadratic in $\beta$ and $D^2(x)$ is the Hessian of $x(t)$ with respect to the entries of $\beta$, that is
\begin{equation}\label{hessian}
	D^2(x) = \left[ \begin{matrix} D_{x(0)}^2(x(t)) & D_{x(0),\alpha}^2(x(t)) \\
			D_{\alpha, x(0)}^2(x(t)) & D_{\alpha}^2(x(t)) \\
	\end{matrix} \right]
\end{equation}
where $D_{a}^2(g) = \frac{\partial^2 g}{\partial a^2}$ and $D_{a,b}^2(g) = \frac{\partial^2 g}{\partial a \partial b}$.  The second partial derivatives of $g$ are also called the second-order sensitivities of $g$ with respect to $a$ and $b$.  Similarly, from \eqref{hxtaylor},
\begin{equation}\label{dh}
	\Delta h = \delta h + \delta^2 h = D_x(h) \Delta x + \frac{1}{2} D_x^2(h) (\Delta x)^2
\end{equation}
Substituting \eqref{dxx} in \eqref{dh} and using \eqref{dherr} for the case of the single observation, we obtain a nonlinear least-squares problem.
\begin{equation}\label{sols}
	e_t = D_x(h)(\dx + \dxx) + \frac{1}{2}D_x^2(h)(\dx+\dxx)^2
\end{equation}
The unknown $\beta$ is then obtained by minimizing 
\begin{equation}\label{sofunctional}
	g_2(\beta) = \left\| e_t - D_x(h)(\dx + \dxx) + \frac{1}{2}D_x^2(h)(\dx+\dxx)^2 \right\|^2
\end{equation}
In the special case where the state is directly observable, $h(x) = x$ and so $D_x(h) = 1$ and $D_x^2(x) = 0$.  Substituting these into \eqref{sofunctional}, in view of \eqref{derxmatrix} and \eqref{hi}, we get
\begin{equation}\label{hidentity}
	g_2(\beta) = \left\| e_t - \overline{H}\beta - \frac{1}{2}\beta^\T D^2(x) \beta\right\|^2
\end{equation}
Clearly, $g_2(\beta)$ is a fourth degree polynomial in the components of $\beta$ and in general can have multiple minima which could further complicate the minimization problem. When there are $N$ $(\geq 2)$ observations at times $0 \leq t_0 < t_1 < \cdots < t_{N-1}$, then we get $N$ versions of the relation \eqref{sols}, one for each time. The unknown $\beta$ is then obtained by minimizing
\begin{equation}\label{sec_ord_multi}
	G_2(\beta) = \sum_{i = 0}^{N-1} \left\| e_{t_i} - D_{x(t_i)}(h)[\dx(t_i)+\delta^2x(t_i)]-\frac{1}{2}D_{x(t_i)}^2(h)[\dx(t_i)+\delta^2x(t_i)]^2\right\|^2.
\end{equation}
In the special case when $h(x) = x$, $D_x(h) = 1$ and $D_x^2(x) = 0$ and \eqref{sec_ord_multi} reduces to 
\begin{equation}\label{sec_ord_multi_hid}
	G_2(\beta) = \sum_{i = 0}^{N-1} \left\| e_{t_i} - \overline{H}(t_i)\beta-\frac{1}{2}\beta^\T D^2(x(t_i) \beta \right\|^2.
\end{equation}
\\
\section{The Logistic Equation}
Consider the standard logistic equation that describes the growth of a certain population in an environment with carrying capacity 1 and growth rate parameter $\alpha > 0$ given by
\begin{equation}\label{logmodel}
	\dot{x} = \alpha x(1-x)
\end{equation}
with $x(0) = x_o > 0$ as the initial condition.  The solution of this nonlinear ordinary differential equation is given by 
\begin{equation}\label{logsoln}
	x(t) = \frac{x_o e^{\alpha t}}{1-x_o+x_o e^{\alpha t}}
\end{equation}
Notice that $x(t)$ depends nonlinearly on $x_o$ and $\alpha$.

It can be verified that $x(t) \rightarrow 1$ as $t \rightarrow \infty$ and $x(t) \in [0,1]$ for all $t$ whenever $x(0) \in [0,1]$, for $\alpha > 0$.  Since $\dot{x}(t) > 0$ for all $t$, and so is an increasing function which increases from $x_o$ to $1$ as $t$ increases from $0$ to $\infty$.  A typical plot of $x(t)$ in \eqref{logsoln} for $x(0) = 0.5$ and $\alpha = 1$ is given in Figure 1.

The various first-order sensitivities of the the solution with respect to $\beta = (x(0),\alpha)^\T$ are given by
\begin{align}
D_{x(0)}(x) & = \frac{e^{\alpha t}}{[1-x_o+x_o e^{\alpha t}]^2} \label{dxox}\\
D_\alpha(x) &= \frac{x_o(1-x_o)te^{\alpha t}}{[1-x_o+x_o e^{\alpha t}]^2} = x_o(1-x_o)tD_{x(0)}(x) \label{dax}
\end{align}
That is, $D_\alpha(x)$ is a multiple of $D_{x(0)}$.  Further it can be verified for large $t$ that
\begin{equation*}\label{longtermsens}
 D_\alpha(x) \approx \frac{1-x_o}{x_o} te^{-\alpha t} \mbox{ and } D_{x(0)}(x) \approx \frac{1}{{x_o}^2}e^{-\alpha t}
\end{equation*}
Thus, $D_\alpha(x)$ and hence $D_{x(0)}(x)$ tend to zero as $t \rightarrow \infty$.  A plot of $D_\alpha(t)$ and $D_{x(0)}(x)$ versus $t$ for $x_o = 0.5$ and $\alpha = 1$ is given in Figure 2.  This in turn implies that $x(t)$ becomes less sensitive with respect to $\alpha$ and $x(0)$ for large $t$, and that for our framework to be effective, the observations have to be taken during the transient phase of the solution.

The Hessian of $x(t)$ with respect to $\beta$ whose elements are the second-order sensitivity functions are given by
\begin{equation}\label{sosensfcns}
D^2(x) = \left[\begin{matrix} D_{x(0)}^2(x) & D_{x(0),\alpha}^2(x) \\ D_{\alpha,x(0)}^2(x) & D_{\alpha}^2(x) \end{matrix}\right]
\end{equation}
where 
\begin{equation*}\label{d2x0}
D_{x(0)}^2(x) = \frac{2(e^{\alpha t}-e^{2 \alpha t})}{[1-x_o + x_o e^{\alpha t}]^3}
\end{equation*}
\begin{equation*}\label{d2a}
D_{\alpha}^2(x) = \frac{x_o(1-x_o)t^2e^{\alpha t}[1-x_o-x_o e^{\alpha t}]}{[1-x_o+x_o e^{\alpha t}]^3}
\end{equation*}
and
\begin{equation*}\label{d2x0a}
D_{x(0),\alpha}^2(x) = D_{\alpha,x(0)}^2(x) = \frac{t^2 e^{\alpha t} [1-x_o-x_o e^{\alpha t}]}{[1-x_o+x_o e^{\alpha t}]^3}.
\end{equation*}
It is assumed that $x(t)$ is directly observable, and so $h(x) = x$.  Accordingly $D_x(h) \equiv 1$ and $D_x^2(h) \equiv 0$.
\subsection{The First-Order Approximation}
In this case \eqref{hbe} reduces to \eqref{derxmatrix} and from \eqref{dxox} and \eqref{dax} we get
\begin{equation}\label{logh}
H=\overline{H}=D_{x(0)}(x)[1,x_0(1-x_0)t] \in \R^{1\mathsf{x}2}.
\end{equation}
Two cases arise. The first case is when there are less observations than parameters.  The second is when the number of observations is at least as great as the number of unknown parameters.  In this paper, we only address the latter case. In this case, $N > 1$ and 
\begin{equation}\label{odsol}
H_N=\left[ \begin{matrix} D_{x(0)}(x(t_1)) & D_{x(0)}(x(t_1))x_o(1-x_o)t_1 \\
D_{x(0)}(x(t_2)) & D_{x(0)}(x(t_2))x_o(1-x_o)t_2 \\
\vdots  & \vdots \\
D_{x(0)}(x(t_N)) & D_{x(0)}(x(t_N))x_o(1-x_o)t_N \end{matrix} \right]
\end{equation}
It can be verified that matrix $H_N \in \R^{N\mathsf{x}2}$ is of rank two and the least-square solution is given by
\begin{equation}\label{odlssol}
\beta_{LS}^f = \left\{ \begin{matrix} H_N^{-1} E_N & if \mbox{ } N = 2\\
(H_N^\T H_N)^{-1}H_N^\T E_N & if \mbox{ } N > 2
\end{matrix} \right.
\end{equation}

\subsection{The Second-Order Approximation}
Expanding the right hand side of \eqref{sec_ord_multi_hid} and dropping the terms of degree 3 and 4, we obtain a quadratic approximation $Q(\beta)$ to $g_2(\beta)$ given by
\begin{equation}\label{quad}
Q(\beta)=\sum_{i = 0}^{N-1}\left[e_{t_i}^2-2e_{t_i}\overline{H}(t_i)\beta+\beta^\T[\overline{H}(t_i)^\T\overline{H}(t_i)-e_{t_i} D^2(x(t_i))]\beta\right].
\end{equation}
The $\beta$ that minimizes this is given by 
\begin{equation}\label{sobls}
\beta_{LS}^s =\sum_{i = 0}^{N-1} \left[\overline{H}(t_i)^\T\overline{H}(t_i) - e_{t_i} D^2(x(t_i))\right]\overline{H}(t_i)^\T e_{t_i}.
\end{equation}

\section{Numerical Experiments}

In this section we compare the performance of the first-order and the second-order methods using $N$ $(\geq 2)$ observations.
\subsection{Generation of Observations}
It is assumed that $h(x) = x$, that is, the state of the system $x(t)$ is directly observable and that $z_t = x(t)$.  We use the model solution $\overline{x}(t)$ starting from $\overline{x}(0)$ and $\overline{\alpha}$ to generate different sets of $N$ observations.

The distribution of the $N$ observation times
\[ t_0 < t_1 < t_2 < \dots < t_{N-1} \]
are given in Table 1 where $t_i = t_0 + i(k\Delta)$ with $\Delta = 0.5$ (fixed), $k$ = 1, 4, 8 and 12 and $0 \leq i \leq N-1$.  Here $t_0$ denotes the starting time and $(k\Delta)$ denotes the time between successive observations.  Thus, for example, for $t_0 = 0$ and $k = 4$, we generate $N = 6$ observations at times $t_0 = 0.0$, $ t_1 = 2.0$, $t_2 = 4.0$, $ t_3 = 6.0$, $t_4 = 8.0$, and $ t_5 = 10.0$, which corresponds to the second row of Table 1.

The actual observations $z_{t_i} = \overline{x}(t_i)$ where $\overline{x}(t_i)$ is the solution of \eqref{logmodel} starting from $\overline{x}(0) = 0.5$ and $\overline{\alpha} = 1.0$ used in the estimation of $\beta$ are given in Table 2.  
\subsection{A Plan for the Experiment}
Let $x(t)$ be the solution of the model \eqref{logmodel} starting from $x(0) \neq \overline{x}(0)$ and $\alpha \neq \overline{\alpha}$.  Our goal is to estimate $\dx(0) = x(0)-\overline{x}(0)$ and $\da = \alpha - \overline{\alpha}$ using the set of observations in Table 2.  We divide the experiment into  three parts based on $N$, the number of observations.
\\
\underline{Case 1}: $N = 2$.  We compute the estimates of $\beta$ using two observations at times $t_0$ and $t_0 + k\Delta$ for $k$ = 1, 4, 8 and 12 and for $t_0$ = 0, 4 and 8.  Estimates of $\beta$ using the first-order and the second-order methods are given in the top part of Tables 3 and 4 respectively.

For example, referring to the top part of Table 3, the estimate of $\beta = (\dx(0) \mbox{ } \da)$ using two observations $z_{t_0} = z_4 = x(4)$ and $z_{t_1} = z_{4.5} = x(4.5)$ and the first-order method is given by $\beta = (-0.0955, 0.0981)$.  Since the actual value of $\dx(0) = -0.1$ and $\da = 0.1$, the relative error in the estimate of $\dx(0) = \frac{-0.0955-(-0.1)}{-0.1} = 0.045$ and that of $\da = \frac{0.0981-0.1}{0.1} = -0.019$.  Similarly for all other patterns of $N = 2$ observations.
\\
\underline{Case 2}: $N = 4$. Results of the estimation of $\beta$ using 4 observations at times $t_0$, $t_0 + k\Delta$, $t_0+2k\Delta$ and $t_0+3k\Delta$ for $k$ = 1, 4, 8 and 12 and $t_0$ = 0, 4 and 8 are given in the middle parts of Tables 3 and 4.
\\
\underline{Case 3}: $N = 6$. Results of the estimation of $\beta$ using 4 observations at times $t_0$, $t_0 + k\Delta$, $t_0+2k\Delta$, ...,  $t_0+5k\Delta$ for $k$ = 1, 4, 8 and 12 and $t_0$ = 0, 4 and 8 are given in the bottom parts of Tables 3 and 4.
 \\
 \subsection{Comments and Discussion}
 A number of observations are in order.
 \begin{enumerate}
 \item When two observations are too close to each other, say by a distance $\Delta = 0.5$, the solutions at these two times are very close to each other and consequently the sensitivity functions at time $t_0$ $(D_{x(0)}(x(t_0))$, $D_\alpha(x(t_0)))$ and at time $t_1 = t_0+\Delta$ namely $(D_{x(0)}(x(t_1))$, $D_\alpha(x(t_1)))$ are also very close to each other.  This in turn implies that the two rows in $\overline{H}_N$ are very nearly colinear, which leads to bad conditioning of the matrix ${\overline{H}_N}^\T \overline{H}_N$.  Hence there is a greater chance for the estimates to be less reliable.
\item From \eqref{logsoln} and Figure 1, it follows that the solution $x(t)$ attains the steady state value which is independent of $x(0)$ and $\alpha$.  In other words, the sensitivity of $x(t)$ with respect to $x(0)$ and $\alpha$ decrease to zero as $t$ increases.  Thus, from \eqref{dxox} and \eqref{dax} it follows that  $D_{x(0)}(x(t))$ and $D_\alpha(x(t))$ both tend to zero as $t$ increases.  Thus, the row of $\overline{H}_N$ for large $t$ becomes zero leading to ill-conditioning of $\overline{H}_N$ which makes the estimates less reliable.
\item Since the solution $x(t)$ depends on $x(0)$ and $\alpha$ nonlinearly, we cannot hope to recover the actual perturbations $\dx_0$ and $\da$ that led to the forecast bias in the first place.  While in principle the second-order method is better than the first-order counterpart, to improve the overall accuracy of the estimate, one may have to resort to an iterative improvement of the estimate.  A detailed comparison of the performance of the iterative versions of the first and second order methods for small values of the perturbations ($\dx_0 = -0.1$ and $\da = 0.1$) are given in Tables 5 and 6 respectively.  Tables 7 and 8 contain similar results for large values of $\dx_0$ and $\da$.

It turns out that the iterative versions of the first-order method peform at least as well as the iterative versions of the second-order counterpart.  Given that second-order methods require larger computation, from this exercise it follows that the iterative first-order method would be a good choice for estimating the forecast bias.
\end{enumerate}
 \newpage
\section{References}
Cacuci, D.G. (2003) \textit{Sensitivity and Uncertainty Analysis}, Chapman and Hall/CRC Press, Boca Raton, FL\\
\\
Cruz, Jr. J.B. (1973) \textit{System Sensitivity Analysis}, Stroudsberg, PA. (editor)\\
\\
Lakshmivarahan, S. and J. Lewis (2008) "Finding sources of bias error in forecast models: A framework," \textit{Monthly Weather Review}. (submitted)\\
\\
Lewis, J., S. Lakshmivarahan, and S.K. Dhall (2006) \textit{Dynamic Data Assimilation}, Cambridge University Press, 654 pages\\
\newpage
\begin{figure}[c]
\center
\includegraphics[width=70mm,height=60mm]{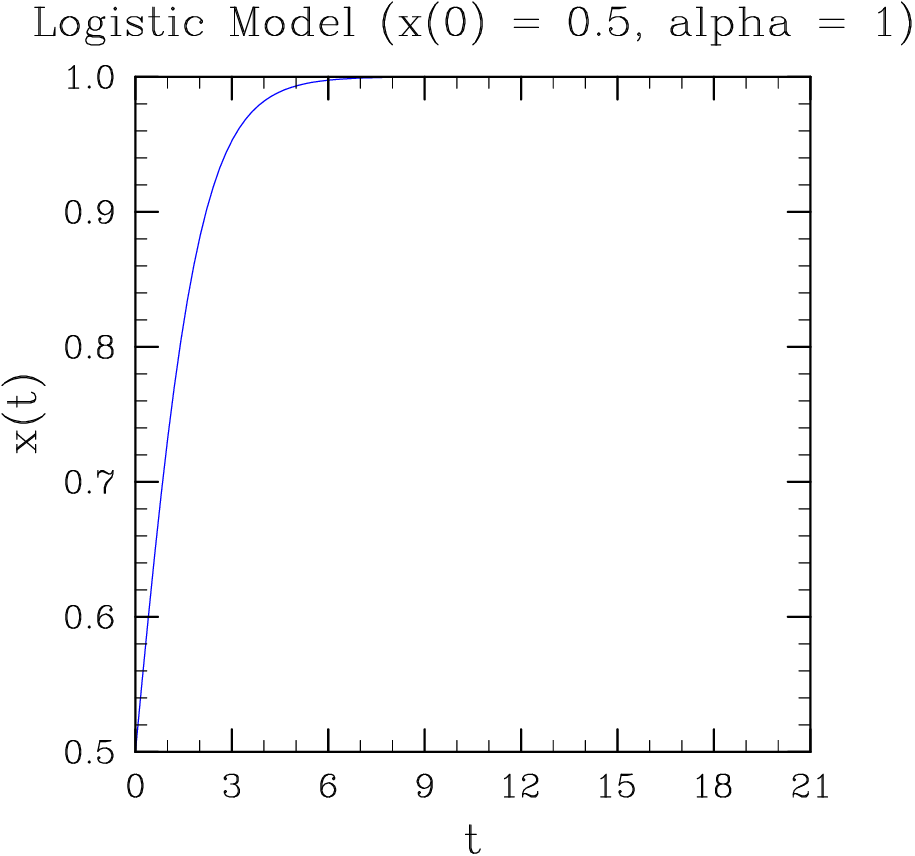}
\caption{Typical Solution of Logistic Equation ($x(0) = 0.5$ and $\alpha = 1$)}
\end{figure}
\begin{figure}[h]
\center
\subfigure[Sensitivity with respect to x(0)]{\includegraphics[width=70mm,height=60mm]{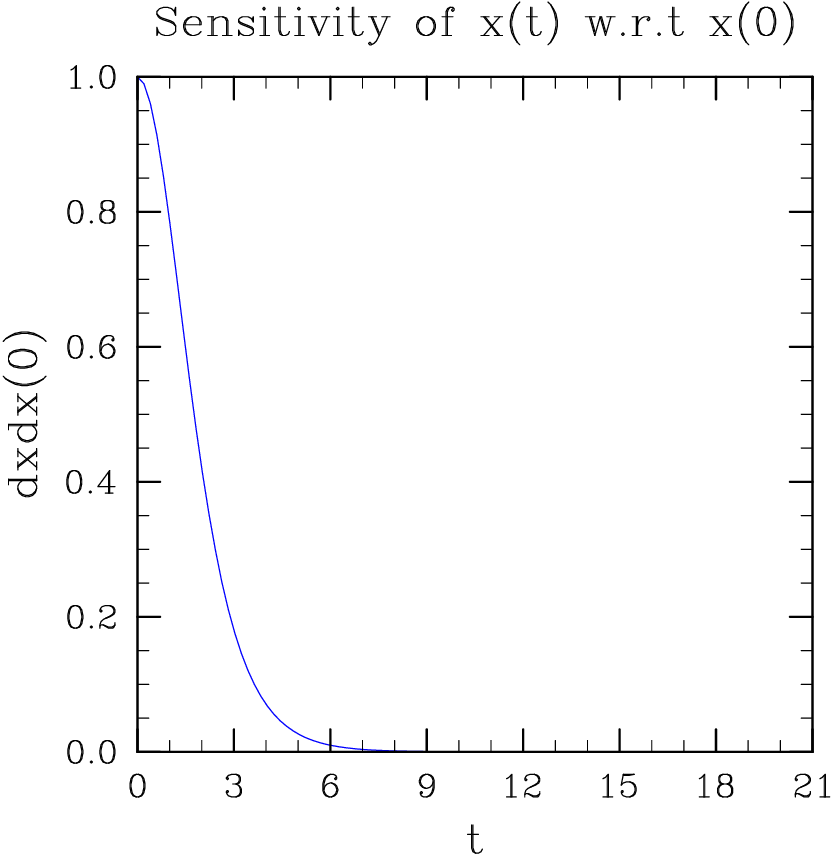}}
\subfigure[Sensitivity with respect to $\alpha$]{\includegraphics[width=70mm,height=60mm]{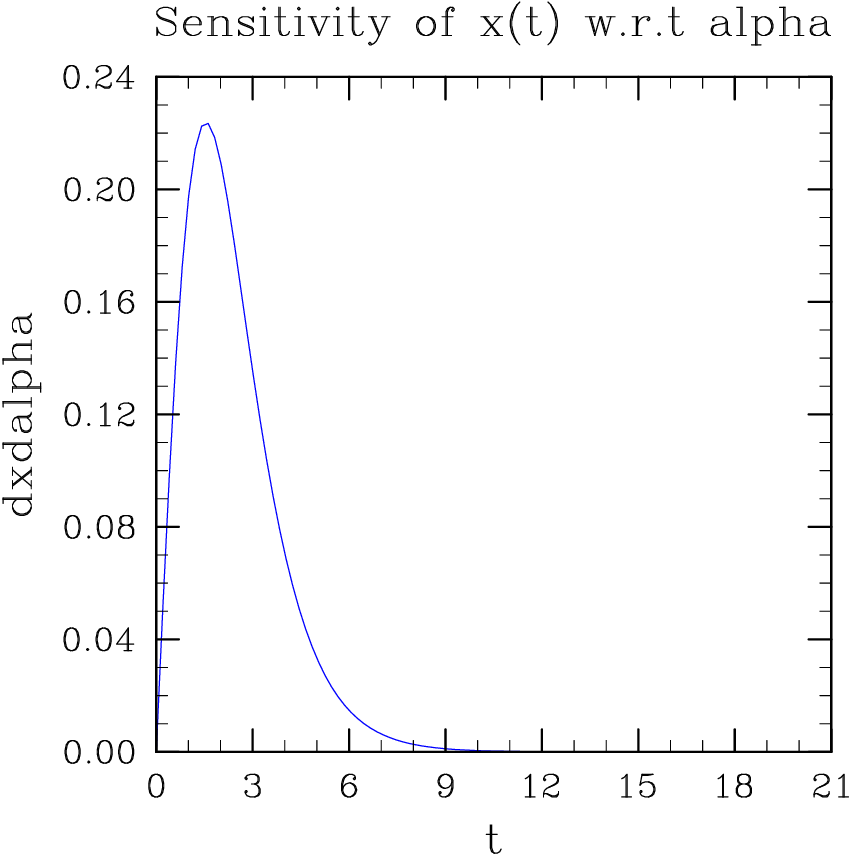}}
\caption{First-Order Sensitivities of the Solution to the Logistic Equation}
\end{figure}

\begin{table}[c]
\center
\begin{tabular}{|c|cccc|}
\hline
\multicolumn{5}{|c|}{Observation Times}\\
\hline
$k\Delta$ & $\Delta$ & 4$\Delta$ & 8$\Delta$ & 12$\Delta$\\
\hline
$t_0$	 &  0.0	 &  0.0	 &  0.0	 &  0.0\\
$t_1$	 &  0.5	 &  2.0	 &  4.0	 &  6.0\\
$t_2$	 &  1.0	 &  4.0	 &  8.0	 & 12.0\\
$t_3$	 &  1.5	 &  6.0	 & 12.0	 & 18.0\\
$t_4$	 &  2.0	 &  8.0	 & 16.0	 & 24.0\\
$t_5$	 &  2.5	 & 10.0	 & 20.0	 & 30.0\\
\hline
$k\Delta$ & $\Delta$ & 4$\Delta$ & 8$\Delta$ & 12$\Delta$\\
\hline
$t_0$	 &  4.0	 &  4.0	 &  4.0	 &  4.0\\
$t_1$	 &  4.5	 &  6.0	 &  8.0	 & 10.0\\
$t_2$	 &  5.0	 &  8.0	 & 12.0	 & 16.0\\
$t_3$	 &  5.5	 & 10.0	 & 16.0	 & 22.0\\
$t_4$	 &  6.0	 & 12.0	 & 20.0	 & 28.0\\
$t_5$	 &  6.5	 & 14.0	 & 24.0	 & 34.0\\ 
\hline
$k\Delta$ & $\Delta$ & 4$\Delta$ & 8$\Delta$ & 12$\Delta$\\
\hline
$t_0$	 &  8.0	 &  8.0	 &  8.0	 &  8.0\\
$t_1$	 &  8.5	 & 10.0	 & 12.0	 & 14.0\\
$t_2$	 &  9.0	 & 12.0	 & 16.0	 & 20.0\\
$t_3$	 &  9.5	 & 14.0	 & 20.0	 & 26.0\\
$t_4$	 & 10.0	 & 16.0	 & 24.0	 & 32.0\\
$t_5$	 & 10.5	 & 18.0	 & 28.0	 & 38.0\\
\hline
\end{tabular}
\caption{Distribution of the $N$ observation times: $t_i = t_0+k(i-1)\Delta$ with $\Delta = 0.5$, $k = 1, 4, 8,$ and $12$ and $0\leq i\leq N-1$.  $t_0$ denotes the starting time and $k\Delta$ denotes the time interval between successive observations and $N$ denotes the number of observations.}
\end{table}
\begin{table}[c]
\center
\begin{tabular}{|c|cccc|}
\hline
\multicolumn{5}{|c|}{Actual Observations}\\
\hline
&\multicolumn{4}{|c|}{Multiple of $\Delta$}\\
\hline
$t_0 = 0$	 &  1	 &  4	 &  8	 & 12\\
\hline
$z_0$	 & 0.50000000000000	 & 0.50000000000000	 & 0.50000000000000	 & 0.50000000000000\\
$z_1$	 & 0.62245933120185	 & 0.88079707797788	 & 0.98201379003791	 & 0.99752737684337\\
$z_2$	 & 0.73105857863000	 & 0.98201379003791	 & 0.99966464986953	 & 0.99999385582540\\
$z_3$	 & 0.81757447619364	 & 0.99752737684337	 & 0.99999385582540	 & 0.99999998477002\\
$z_4$	 & 0.88079707797788	 & 0.99966464986953	 & 0.99999988746484	 & 0.99999999996225\\
$z_5$	 & 0.92414181997876	 & 0.99995460213130	 & 0.99999999793885	 & 0.99999999999991\\
\hline
&\multicolumn{4}{|c|}{Multiple of $\Delta$}\\
\hline
$t_0 = 4$	 &  1	 &  4	 &  8	 & 12\\
\hline
$z_0$	 & 0.98201379003791	 & 0.98201379003791	 & 0.98201379003791	 & 0.98201379003791\\
$z_1$	 & 0.98901305736941	 & 0.99752737684337	 & 0.99966464986953	 & 0.99995460213130\\
$z_2$	 & 0.99330714907572	 & 0.99966464986953	 & 0.99999385582540	 & 0.99999988746484\\
$z_3$	 & 0.99592986228410	 & 0.99995460213130	 & 0.99999988746484	 & 0.99999999972105\\
$z_4$	 & 0.99752737684337	 & 0.99999385582540	 & 0.99999999793885	 & 0.99999999999931\\
$z_5$	 & 0.99849881774326	 & 0.99999916847197	 & 0.99999999996225	 & 1.00000000000000\\
\hline
&\multicolumn{4}{|c|}{Multiple of $\Delta$}\\
\hline
$t_0 = 8$	 &  1	 &  4	 &  8	 & 12\\
\hline
$z_0$	 & 0.99966464986953	 & 0.99966464986953	 & 0.99966464986953	 & 0.99966464986953\\
$z_1$	 & 0.99979657302194	 & 0.99995460213130	 & 0.99999385582540	 & 0.99999916847197\\
$z_2$	 & 0.99987660542401	 & 0.99999385582540	 & 0.99999988746484	 & 0.99999999793885\\
$z_3$	 & 0.99992515377249	 & 0.99999916847197	 & 0.99999999793885	 & 0.99999999999489\\
$z_4$	 & 0.99995460213130	 & 0.99999988746484	 & 0.99999999996225	 & 0.99999999999999\\
$z_5$	 & 0.99997246430889	 & 0.99999998477002	 & 0.99999999999931	 & 1.00000000000000\\
\hline
\end{tabular}
\caption{Actual Observations}
\end{table}

\begin{table}[h]
\center
\begin{tabular}{|c c c c c c|}
\hline
\multicolumn{2}{|c}{True Corrections}&\multicolumn{2}{c}{$\dx_o = -0.1$} &\multicolumn{2}{c|}{$\da = 0.1$}\\
\hline
\multicolumn{2}{|c}{2 observations}& \multicolumn{4}{c| }{multiple of $\Delta$}\\
\hline
$t_o$& &1&4&8&12 \\
\hline
0&$\dx_o$&-0.1000&-0.1000&-0.1000&-0.1000\\
 &$\da$&0.0761&0.0968&0.1028&0.0989\\
 &$\kappa$&2.8e+2&4.1e+1&2.0e+2&3.1e+3\\
 \hline
4&$\dx_o$&-0.0955&-0.0888&-0.0809&-0.0739\\
 &$\da$&0.0981&0.0912&0.0829&0.0756\\
 &$\kappa$&1.3e+3&6.0e+2&5.0e+3&8.1e+4\\
  \hline
8&$\dx_o$&-0.0479&-0.0390&-0.0284&-0.0190\\
 &$\da$&0.0657&0.0611&0.0555&0.0507\\
 &$\kappa$&7.9e+3&3.8e+3&3.2e+4&5.2e+5\\
 \hline
\hline
\multicolumn{2}{|c}{4 observations }& \multicolumn{4}{c| }{multiple of $\Delta$}\\
\hline
$t_o$& &1&4&8&12 \\
\hline
0&$\dx_o$&-0.1006&-0.1000&-0.1000&-0.1000\\
 &$\da$&0.0892&0.0977&0.1028&0.0989\\
 &$\kappa$&7.0e+1&3.5e+1&2.0e+2&3.1e+3\\
 \hline
4&$\dx_o$&-0.1038&-0.1108&-0.1477&0.4009\\
 &$\da$&0.0977&0.1039&0.1349&-0.0570\\
 &$\kappa$&3.9e+2&9.4e+2&5.0e+4&1.5e+5\\
  \hline
8&$\dx_o$&-0.0440&-0.0378&-0.0283&-0.0190\\
 &$\da$&0.0637&0.0605&0.0555&0.0507\\
 &$\kappa$&2.5e+3&3.4e+3&3.2e+4&5.2e+5\\
 \hline
\hline
\multicolumn{2}{|c}{6 observations}& \multicolumn{4}{c| }{multiple of $\Delta$}\\
\hline
$t_o$& &1&4&8&12 \\
\hline
0&$\dx_o$&-0.1011&-0.1000&-0.1000&-0.1000\\
 &$\da$&0.0951&0.0977&0.1028&0.0989\\
 &$\kappa$&4.2e+1&3.5e+1&2.0e+2&3.1e+3\\
 \hline
4&$\dx_o$&-0.0907&-0.0879&-0.0808&-0.0739\\
 &$\da$&0.0932&0.0902&0.0828&0.0756\\
 &$\kappa$&3.0e+2&5.4e+2&5.0e+3&8.1e+4\\
  \hline
8&$\dx_o$&-0.0416&-0.0378&-0.0283&-0.0190\\
 &$\da$&0.0625&0.0605&0.0555&0.0507\\
 &$\kappa$&1.8e+3&3.4e+3&3.2e+4&5.2e+5\\
 \hline
\end{tabular}
 \caption{Estimates of $\beta$ using the first-order method. $\kappa$ denotes the condition number of the matrix $\overline{H}_N^\T \overline{H}_N$, where $N$ is the number of observations used in the estimation.}
\end{table}
\begin{table}
\center
\begin{tabular}{|c c c c c c|}
\hline
\multicolumn{2}{|c}{True Corrections}&\multicolumn{2}{c}{$\dx_o = -0.1$} &\multicolumn{2}{c|}{$\da = 0.1$}\\
\hline
\multicolumn{2}{|c}{2 observations}& \multicolumn{4}{c| }{multiple of $\Delta$}\\
\hline
$t_o$& &1&4&8&12 \\
\hline
0&$\dx_o$&-0.1000&-0.1000&-0.1000&-0.1000\\
 &$\da$&0.1117&0.0987&0.0945&0.0980\\
 &$\kappa$&2.1e+2&2.5e+1&1.8e+2&7.1e+3\\
 \hline
4&$\dx_o$&-0.1009&-0.1098&-0.1489&-0.1264\\
 &$\da$&0.0953&0.1030&0.1359&0.1247\\
 &$\kappa$&1.1e+3&9.6e+2&5.1e+4&2.5e+5\\
  \hline
8&$\dx_o$&0.3763&0.3744&0.3737&0.3749\\
 &$\da$&-0.0528&-0.0523&-0.0521&-0.0524\\
 &$\kappa$&8.2e+3&4.7e+3&5.2e+4&1.1e+6\\
 \hline
\hline
\multicolumn{2}{|c}{4 observations }& \multicolumn{4}{c| }{multiple of $\Delta$}\\
\hline
$t_o$& &1&4&8&12 \\
\hline
0&$\dx_o$&-0.0995&-0.1000&-0.1000&-0.1000\\
 &$\da$&0.1025&0.0983&0.0945&0.0980\\
 &$\kappa$&5.4e+1&2.3e+1&1.8e+2&7.1e+3\\
 \hline
4&$\dx_o$&-0.1038&-0.1108&-0.1477&0.4009\\
 &$\da$&0.0977&0.1039&0.1349&-0.0570\\
 &$\kappa$&3.9e+2&9.4e+2&5.0e+4&1.5e+5\\
  \hline
8&$\dx_o$&0.3754&0.3743&0.3737&0.3749\\
 &$\da$&-0.0525&-0.0523&-0.0521&-0.0524\\
 &$\kappa$&2.6e+3&4.3e+3&5.2e+4&1.1e+6\\
 \hline
\hline
\multicolumn{2}{|c}{6 observations}& \multicolumn{4}{c| }{multiple of $\Delta$}\\
\hline
$t_o$& &1&4&8&12 \\
\hline
0&$\dx_o$&-0.0991&-0.1000&-0.1000&-0.1000\\
 &$\da$&0.0991&0.0983&0.0945&0.0980\\
 &$\kappa$&3.3e+1&2.3e+1&1.8e+2&7.1e+3\\
 \hline
4&$\dx_o$&-0.1059&-0.1108&-0.1477&0.4009\\
 &$\da$&0.0995&0.1039&0.1349&-0.0570\\
 &$\kappa$&3.0e+2&9.4e+2&5.0e+4&1.5e+5\\
  \hline
8&$\dx_o$&0.3750&0.3743&0.3737&0.3749\\
 &$\da$&-0.0524&-0.0523&-0.0521&-0.0524\\
 &$\kappa$&1.9e+3&4.3e+3&5.2e+4&1.1e+6\\
 \hline
\end{tabular}
 \caption{Estimates of $\beta$ using the second-order method. $\kappa$ denotes the condition number of the matrix $\overline{H}_N^\T \overline{H}_N$, where $N$ is the number of observations used in the estimation.}
\end{table}

\begin{table}[h]
\center
\footnotesize
\begin{tabular}{|c c c c c c|}
\hline
\multicolumn{2}{|c}{True Corrections}&\multicolumn{2}{c}{$\dx_o = -0.1$} &\multicolumn{2}{c|}{$\da = 0.1$}\\
\hline
\multicolumn{2}{|c}{2 observations}& \multicolumn{4}{c| }{multiple of $\Delta$}\\
\hline
$t_0 $& &1&4&8&12\\
\hline
0&$\dx_0$&-0.1000&-0.1000&-0.1000&-0.1000\\
&$\da$&0.1000&0.1000&0.1000&0.1000\\
&$\kappa$&2.6e+2&3.2e+1&2.0e+2&4.6e+3\\
&iterations&4&4&4&4\\
\hline
4&$\dx_0$&-0.1000&-0.1000&-0.1000&-0.1000\\
&$\da$&0.1000&0.1000&0.1000&0.1000\\
&$\kappa$&1.4e+3&8.7e+2&1.1e+4&2.7e+5\\
&iterations&4&4&4&5\\
\hline
8&$\dx_0$&-0.1000&-0.1000&-0.1000&-0.1000\\
&$\da$&0.1000&0.1000&0.1000&0.1000\\
&$\kappa$&8.6e+3&5.7e+3&7.5e+4&1.8e+6\\
&iterations&5&5&5&5\\
\hline
\hline
\multicolumn{2}{|c}{4 observations }& \multicolumn{4}{c| }{multiple of $\Delta$}\\
\hline
$t_o$& &1&4&8&12 \\
\hline
0&$\dx_0$&-0.1000&-0.1000&-0.1000&-0.1000\\
&$\da$&0.1000&0.1000&0.1000&0.1000\\
&$\kappa$&6.3e+1&2.9e+1&2.0e+2&4.6e+3\\
&iterations&4&4&4&4\\
\hline
4&$\dx_0$&-0.1000&-0.1000&-0.1000&-0.1000\\
&$\da$&0.1000&0.1000&0.1000&0.1000\\
&$\kappa$&4.6e+2&8.1e+2&1.1e+4&2.7e+5\\
&iterations&4&4&4&5\\
\hline
8&$\dx_0$&-0.1000&-0.1000&-0.1000&-0.1000\\
&$\da$&0.1000&0.1000&0.1000&0.1000\\
&$\kappa$&2.9e+3&5.3e+3&7.4e+4&1.8e+6\\
&iterations&5&5&5&5\\
 \hline
\hline
\multicolumn{2}{|c}{6 observations}& \multicolumn{4}{c| }{multiple of $\Delta$}\\
\hline
$t_o$& &1&4&8&12 \\
\hline
0&$\dx_0$&-0.1000&-0.1000&-0.1000&-0.1000\\
&$\da$&0.1000&0.1000&0.1000&0.1000\\
&$\kappa$&3.9e+1&2.9e+1&2.0e+2&4.6e+3\\
&iterations&4&4&4&4\\
\hline
&$\dx_0$&-0.1000&-0.1000&-0.1000&-0.1000\\
&$\da$&0.1000&0.1000&0.1000&0.1000\\
&$\kappa$&3.5e+2&8.1e+2&1.1e+4&2.7e+5\\
&iterations&4&4&4&5\\
\hline
8&$\dx_0$&-0.1000&-0.1000&-0.1000&-0.1000\\
&$\da$&0.1000&0.1000&0.1000&0.1000\\
&$\kappa$&2.2e+3&5.3e+3&7.4e+4&1.8e+6\\
&iterations&5&5&5&5\\
 \hline
\end{tabular}
 \caption{\footnotesize Estimates of $\beta$ using the iterated first-order method. $\kappa$ denotes the condition number of the matrix $\overline{H}_N^\T \overline{H}_N$, where $N$ is the number of observations used in the estimation. The number of iterations is that required to reach a threshold of $10^-6$ (at most 10)}
\end{table}

\begin{table}[h]
\center
\footnotesize
\begin{tabular}{|c c c c c c|}
\hline
\multicolumn{2}{|c}{True Corrections}&\multicolumn{2}{c}{$\dx_o = -0.1$} &\multicolumn{2}{c|}{$\da = 0.1$}\\
\hline
\multicolumn{2}{|c}{2 observations}& \multicolumn{4}{c| }{multiple of $\Delta$}\\
\hline
$t_0 $& &1&4&8&12\\
\hline
0&$\dx_0$&-0.1000&-0.1000&-0.1000&-0.1000\\
&$\da$&0.1000&0.1000&0.1000&0.1000\\
&$\kappa$&2.6e+2&3.2e+1&2.0e+2&4.6e+3\\
&iterations&4&4&4&4\\
4&$\dx_0$&-0.1000&-0.1000&-0.1000&-0.1000\\
&$\da$&0.1000&0.1000&0.1000&0.1000\\
&$\kappa$&1.4e+3&8.7e+2&1.1e+4&2.7e+5\\
&iterations&4&4&5&6\\
8&$\dx_0$&-0.1000&-0.1000&-0.1000&NaN\\
&$\da$&0.1000&0.1000&0.1000&NaN\\
&$\kappa$&8.6e+3&5.7e+3&7.5e+4&1.0e+9\\
&iterations&6&5&5&10\\
\hline
\hline
\multicolumn{2}{|c}{4 observations }& \multicolumn{4}{c| }{multiple of $\Delta$}\\
\hline
$t_o$& &1&4&8&12 \\
\hline
0&$\dx_0$&-0.1000&-0.1000&-0.1000&-0.1000\\
&$\da$&0.1000&0.1000&0.1000&0.1000\\
&$\kappa$&6.3e+1&2.9e+1&2.0e+2&4.6e+3\\
&iterations&4&4&4&4\\
4&$\dx_0$&-0.1000&-0.1000&-0.1000&-0.0868\\
&$\da$&0.1000&0.1000&0.1000&0.1023\\
&$\kappa$&4.6e+2&8.1e+2&1.1e+4&2.4e+5\\
&iterations&4&4&5&10\\
8&$\dx_0$&-0.1000&-0.1000&-0.1000&-0.1000\\
&$\da$&0.1000&0.1000&0.1000&0.1000\\
&$\kappa$&2.9e+3&5.3e+3&7.4e+4&1.8e+6\\
&iterations&7&7&6&6\\
 \hline
\hline
\multicolumn{2}{|c}{6 observations}& \multicolumn{4}{c| }{multiple of $\Delta$}\\
\hline
$t_o$& &1&4&8&12 \\
\hline
0&$\dx_0$&-0.1000&-0.1000&-0.1000&-0.1000\\
&$\da$&0.1000&0.1000&0.1000&0.1000\\
&$\kappa$&3.9e+1&2.9e+1&2.0e+2&4.6e+3\\
&iterations&3&4&4&4\\
4&$\dx_0$&-0.1000&-0.1000&-0.1000&NaN\\
&$\da$&0.1000&0.1000&0.1000&NaN\\
&$\kappa$&3.5e+2&8.1e+2&1.1e+4&1.0e+12\\
&iterations&4&4&5&10\\
8&$\dx_0$&-0.1000&-0.1000&-0.1000&NaN\\
&$\da$&0.1000&0.1000&0.1000&NaN\\
&$\kappa$&2.2e+3&5.3e+3&7.4e+4&7.2e+12\\
&iterations&7&7&6&10\\
 \hline
\end{tabular}
 \caption{\footnotesize Estimates of $\beta$ using the iterated second-order method. $\kappa$ denotes the condition number of the matrix $\overline{H}_N^\T \overline{H}_N$, where $N$ is the number of observations used in the estimation. The number of iterations is that required to reach a threshold of $10^-6$ (at most 10)}
\end{table}

\begin{table}[h]
\center
\footnotesize
\begin{tabular}{|c c c c c c|}
\hline
\multicolumn{6}{|c|}{True Parameter Values: $x_o = 0.5$, $\alpha = 1.0$}\\
\hline
&\multicolumn{2}{c|}{Model Values}&\multicolumn{2}{c}{Corrections}&\\
\hline
$t_0$&$x_0$&$\alpha$&$\dx_0$&$\da$&Iterations\\
\hline
0&0.3&   0.8& 0.2000& 0.2000&  5\\
&0.3&   0.9& 0.2000& 0.1000&  5\\
 &0.3&   1.1& 0.2000&-0.1000&  5\\
 &0.3&   1.2& 0.2000&-0.2000&  5\\
 &  0.4&   0.8& 0.1000& 0.2000&  5\\
 &0.4&   1.1& 0.1000&-0.1000&  4\\
 & 0.4&   0.9& 0.1000& 0.1000&  4\\
  & 0.4&   1.2& 0.1000&-0.2000&  4\\
&0.6&   0.8&-0.1000& 0.2000&  4\\
  &0.6&   0.9&-0.1000& 0.1000&  4\\
  &0.6 &  1.1&-0.1000&-0.1000& 5\\
 & 0.6&   1.2&-0.1000&-0.2000&  5\\
 &0.7&   0.8&-0.2000& 0.2000&  4\\
  &0.7&   0.9&-0.2000& 0.1000&  5\\
 &0.7&   1.1&-0.2000&-0.1000&  5\\
 &0.7&   1.2&-0.2000&-0.2000&  5\\
   \hline
4 &0.3&   0.8& 0.2000& 0.2000&  7\\
& 0.3&   0.9& 0.2000& 0.1000&  6\\
 &0.3&   1.1& 0.2000&-0.1000&  5\\
 &0.3&   1.2& 0.2000&-0.2000&  5\\
 &0.4&   1.1& 0.1000&-0.1000&  4\\
& 0.4&   1.2& 0.1000&-0.2000&  5\\
 & 0.4&   0.8& 0.1000& 0.2000&  6\\
 &0.4&   0.9& 0.1000& 0.1000&  5\\
 &0.6&   0.8&-0.1000& 0.2000&  5\\
 &0.6&   0.9&-0.1000& 0.1000&  4\\
&0.6 &  1.1&-0.1000&-0.1000& 6\\
& 0.6&   1.2&-0.1000&-0.2000&  8\\
 &0.7&   0.8&-0.2000& 0.2000&  5\\
 &0.7&   0.9&-0.2000& 0.1000&  5\\
 & 0.7&   1.1&-0.2000&-0.1000&  7\\
 & 0.7&   1.2&-0.2000&-0.2000& 10\\
\hline
8  &0.3&   0.8& 0.2000& 0.2000&  7\\
&0.3&   1.1& 0.2000&-0.1000&  5\\
 &0.3&   0.9& 0.2000& 0.1000&  7\\
 &0.3&   1.2& 0.2000&-0.2000&  7\\
&0.4&   0.8& 0.1000& 0.2000&  7\\
  &0.4&   0.9& 0.1000& 0.1000&  6\\
  &0.4&   1.1& 0.1000&-0.1000&  5\\
&0.4&   1.2& 0.1000&-0.2000& 11\\
&0.6&   0.8&-0.1000& 0.2000&  6\\
&0.6 &  1.1&-0.1000&-0.1000& 6\\
&0.6&   1.2&    NaN&    NaN&100\\
 &0.6&   0.9&-0.1000& 0.1000&  5\\
 &0.7&   1.1&-0.2000&-0.1000&  9\\
 &0.7&   1.2&    NaN&    NaN&100\\
 &0.7&   0.8&-0.2000& 0.2000&  6\\
 &0.7&   0.9&-0.2000& 0.1000&  5\\
\hline
\end{tabular}
\caption{Iterated First-Order Corrections Using Different Model Parameters $\alpha$ and Initial Conditions $x_0$, in all cases using 4 observations spaced 0.5 seconds apart. }
\end{table}
\newpage
\cfoot{ }
\chead{\thepage}
\begin{table}[h]
\center
\footnotesize
\begin{tabular}{|c c c c c c|}
\hline
\multicolumn{6}{|c|}{True Parameter Values: $x_o = 0.5$, $\alpha = 1.0$}\\
\hline
&\multicolumn{2}{c|}{Model Values}&\multicolumn{2}{c}{Corrections}&\\
\hline
$t_0$&$x_0$&$\alpha$&$\dx_0$&$\da$&Iterations\\
\hline
0& 0.3&   0.8&0.2000&0.2000&  5\\
&  0.3&   0.9&0.2000&0.1000&  5\\
  &0.3&   1.1&0.2000&-0.1000&  5\\ 
& 0.3&   1.2&0.2000&-0.2000&  5\\
 & 0.4&   0.8&0.1000&0.2000&  5\\
 &0.4&   0.9&0.1000&0.1000&  5\\
   &0.4&   1.1&0.1000&-0.1000&  4\\
&0.4&   1.2&0.1000&-0.2000&  4\\
 &0.6&   0.8&-0.1000&0.2000&  4\\
  &0.6&   0.9&-0.1000&0.1000&  4\\
  &0.6 &  1.1&-0.1000&-0.1000& 4\\
 &0.6&   1.2&-0.1000&-0.2000&  4\\
 &0.7&   0.8&-0.2000&0.2000&  4\\
  &0.7&   0.9&-0.2000&0.1000&  4\\
 & 0.7&   1.1&-0.2000&-0.1000&  5\\
 & 0.7&   1.2&-0.2000&-0.2000&  5\\
   \hline
4 &0.3&   0.8&   NaN&   NaN&101\\
   &0.3&   0.9&0.2000&0.1000&  5\\
  &  0.3&   1.1&0.2000&-0.1000&  5\\
  &0.3&   1.2&0.2000&-0.2000&  7\\
  &0.4&   0.8&0.1000&0.2000&  7\\
 &0.4&   0.9&   NaN&   NaN&100\\
 &0.4&   1.1&0.1000&-0.1000&  4\\
  &0.4&   1.2&0.1000&-0.2000&  5\\
&0.6&   0.8&-0.1000&0.2000&  7\\
  &0.6&   0.9&-0.1000&0.1000&  4\\
 &0.6  & 1.1&-0.1000&-0.1000& 5\\
 & 0.6&   1.2&-0.1000&-0.2000&  5\\
 &0.7&   0.8&-0.2000&0.2000&  5\\
 &0.7&   0.9&-0.2000&0.1000&  4\\
 & 0.7&   1.1&-0.2000&-0.1000&  5\\
  &0.7&   1.2&-0.2000&-0.2000&  5\\
    \hline
8 &0.3&   0.8&   NaN&   NaN&100\\
  & 0.3&   1.1&0.2000&-0.1000&  6\\
 &0.3&   1.2&0.2000&-0.2000&  5\\
  &0.3&   0.9&0.2000&0.1000& 16\\
 &0.4&   0.8&   NaN&   NaN&100\\
  &0.4&   0.9&   NaN&   NaN&100\\
 &0.4&   1.1&0.1000&-0.1000&  5\\
&0.4&   1.2&0.1000&-0.2000&  4\\
 &0.6&   0.8&   NaN&   NaN&100\\
  &0.6&   0.9&-0.1000&0.1000&  7\\
  & 0.6  & 1.1&-0.1000&-0.1000& 5\\
 & 0.6&   1.2&-0.1000&-0.2000&  6\\
 &0.7&   0.8&-0.2000&0.2000& 12\\
 &0.7&   0.9&-0.2000&0.1000&  5\\
&0.7&   1.1&-0.2000&-0.1000&  5\\
 &0.7&   1.2&-0.2000&-0.2000&  9\\
\hline
\end{tabular}
\caption{Iterated Second-Order Corrections Using Different Model Parameters $\alpha$ and Initial Conditions $x_0$, in all cases using 4 observations spaced 0.5 seconds apart. }
\end{table}
\end{document}